# Technologies for 3D Wafer Level Heterogeneous Integration

M.J. Wolf*, P. Ramm,** A. Klumpp**, H. Reichl*

Fraunhofer IZM

*Berlin/**Munich

Contact: wolf@izm.fraunhofer.de

*Abstract*-3D integration is a fast growing field that encompasses different types of technologies. The paper addresses one of the most promising technology which uses Through Silicon Vias (TSV) for interconnecting stacked devices on wafer level to perform high density interconnects with a good electrical performance at the smallest form factor for 3D architectures. Fraunhofer IZM has developed a post front-end 3D integration process which allows stacking of functional and tested FE-devices e.g. sensors, ASICs on wafer level as well as a technology portfolio for passive silicon interposer with redistribution layers and TSV.

## I. Drivers for 3D System Integration

Since several years packaging is driven by System in Package (SiP) solutions to meet the requirements of improved performance, miniaturization and cost reduction. This leads to a number of technologies where 3D system integration is one of the main potential drivers [1].

In general, the introduction of 3D integration technologies is driven by

- *Form factor*: Reduction of system volume, weight and footprint
- *Performance*: Improvement of integration density and reduction of interconnect length leading to improved transmission speed and reduced power consumption
- *High volume low cost production*: Reduction of processing costs for, e.g., mixed technologies
- *New applications*: e.g. ultra compact camera and detector systems and small wireless sensor nodes

In competition to Systems on Chip (SoC) solutions, the 3D wafer level system integration enables the combination of different optimized production technologies. In addition, 3D integration is a possible solution to overcome the "wiring crisis" caused by signal propagation delay, both, at board and at chip level, because it allows minimal interconnection lengths and the elimination of speed-limiting intra- and inter-chip interconnects. The introduction of very advanced microelectronic systems, as e.g. 3D image processors, will be mainly driven by the enhancement of performance. The potential for low cost fabrication will be a further key aspect for future applications of 3D integration as well. Today, the fabrication of Systems on Chip (SoC) is based on embedding multiple technologies by monolithic integration. But there are serious disadvantages: The chip partition with the highest complexity drives the process technology which leads to a "cost explosion" of the overall system. In contrast to this, suitable 3D integration technologies enable the combination of different optimized base technologies, e.g. MEMS, CMOS, etc., with the potential of low cost fabrication through high yield and high miniaturization degree.

## II. Advanced 3D Wafer Level System Integration Technologies

### 2.1 Through Silicon Via (TSV) Technology

Wafer level packaging technologies, e.g. CSP with redistribution layers or flip chip mounted devices on wafer, are already introduced in high volume production. Currently, different technologies which use Through Silicon Vias (TSV) in active or passive silicon devices are in development to satisfy the need to increase performance and functionality while reducing size, power and cost of the system. Today, there are two mainstreams to realize TSVs. One is the implementation into the front-end CMOS process and the second is a post front end process (via first/via last) process. Both scenarios have pros and cons and the selection depends on application and infrastructure. The post front end process allows the realization of compact 3D system architectures as a packaging task with complete tested device wafers independent from the device wafer manufacturer.

Key process technologies enabling 3D architectures with TSV interconnects include:

- via formation with high aspect ratio,
- isolation, barrier and seed deposition,
- via metal filling, redistribution lines (RDL),
- wafer thinning,
- thin wafer handling and transfer processes,
- assembly: wafer/chip alignment, adjusted bonding

Most of those 3D technologies are quite new to the packaging industry and require a FE/BE infrastructure. That's why 3D-IC architectures are today still at the R&D stage, even in the largest IC companies, but they are in focus as a potential solution with a high priority. Many of the key technical issues and challenges for TSV interconnects are not fully resolved yet. There are also a number of alternative technologies, e.g.:

- process integration: via-first vs. via-last
- via filling: materials (e.g. poly Si, Cu, W, conductive polymer, metal paste) and techniques (e.g. electroplating, CVD, polymer coating),
- wafer level assembly: chip-to-chip chip-to-wafer or wafer-to-wafer,
- bonding: soldering, direct Cu-Cu, adhesive, direct fusion

The development of selected technical parameters for TSVs is given in Table 1. This data are provided by ITRS and postulates volume production.

**Table 1: Key technical parameters for stacked architectures using TSV**

| Year of Production | 2007 | 2008 | 2009 | 2010 | 2011 | 2012 | 2013 |
|---|---|---|---|---|---|---|---|
| Numbers of stacked die using TSV | 3 – (8) | 6 | 9 | >9 | >9 | >9 | >9 |
| Minimum TSV pitch | 10.0 | 8.0 | 6.0 | 5.0 | 4.0 | 3.8 | 3.6 |
| TSV maximum aspect ratio** | 10.0 | 10.0 | 10.0 | 10.0 | 10.0 | 10.0 | 10.0 |
| TSV exit diameter (µm) | 4.0 | 4.0 | 3.0 | 2.5 | 2.0 | 1.9 | 1.8 |
| TSV layer thickness for minimum pitch | 50 | 20 | 15 | 15 | 10 | 10 | 10 |

. **This applies for small diameter vias. The larger diameter vias will have a smaller aspect ratio.*

### 2.2 SI INTERPOSER AND THIN CHIP INTEGRATION

Silicon Interposer as a carrier substrate for flip chip assembled dies are in special focus for applications which require high density wiring and interconnects e.g. image sensor systems, memory, processor etc.

In combination with Through Silicon Vias, these Si carriers allows the realization of double side flip chip assembled modules with a very small form factor. In [2] an example is given for a silicon interposer with TSV´s for a RF transceiver module. Thin film technology (polymer-copper) with integrated passive devices (R,L,C) was used to realize the wiring for the flip chip assembled transceiver.

The transceiver was solder bumped (SnAg) using ECD. The bottom side of the carrier has IO terminals for larger solder balls (preforms) which provide the interconnection to the printed circuit board. Fig. 1a and Fig. 1b show the wafer level assembled module and a cross-section.

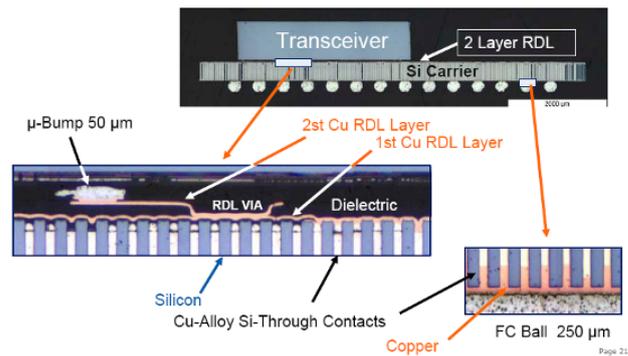

**Fig. 1b Cross-section of RF-Si module [2]**

For the realization of the Through Silicon Vias (TSV) in silicon interposer or silicon device wafers different approaches for the via etching (e.g. wet etching, DRIE or laser drilling) and via metallization (e.g. CVD W, CVD Cu, ECD-Cu doped silicon or metal paste) can be performed. The selection of the technology is determined by system requirements (via size and density, aspect ratio, electrical resistivity etc). Metal filling of TSV using electroplating is especially suited for via sizes between 5 µm and 20 µm, which are in special focus for silicon interposer with TSV as a passive carrier substrate. After DRIE and sidewall isolation, the seed layer can be applied by CVD (e.g. Cu or W) or an adequate sputtering processes, e.g. Ti/W:Cu. Fig. 2 shows a Through Silicon Via (diameter (15 µm) filled with Cu by electro deposition using a thin sputtered TiW/Cu seed layer [3, 4]. Copper will also be deposited on the wafer front side during the via plating, which will be removed by a later etching step. Depending on the via sizes and depth (aspect ratio -ASR), a wafer thinning (grinding, CMP, etching) from the backside is required to get access to the metalized vias. The IO terminals on the backside are realized by standard thin film processing (polymer – Cu) followed by solder ball placement. A mechanical support of the interposer during backside processing can be provided by temporary bonding on a carrier substrate (e.g. silicon).

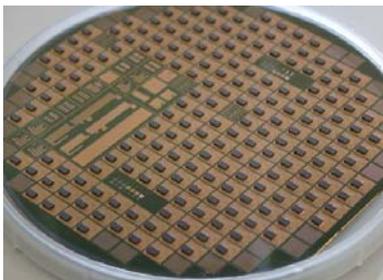

**Fig. 1a WL assembled RF transceiver module with TSV-Si-interposer and integrated passive devices [2]**

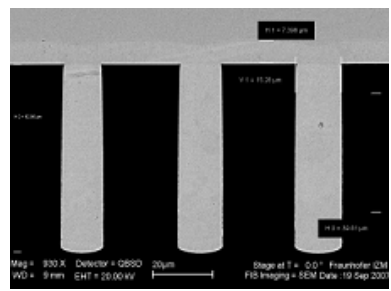

**Fig. 2 ECD-Cu filled TSV (18 µm diameter and 70 µm depth) [3]**

### 2.3 VERTICAL SYSTEM INTEGRATION (VSI) BY ICV-SLID TECHNOLOGY

Vertical System Integration (VSI®) [5] is characterized by bonding and very high density vertical inter-chip wiring of stacked thinned device substrates (Si) with freely positioned Through-Silicon-Vias (TSV) by using standard silicon wafer processes (mainly backend-of-line) (Fig. 3). The VSI-TSV [6] approach can provide the shortest and most plentiful z-axis connections. The TSV technology has various potential benefits:

a) connection lengths can be as short as the thickness of a die, which has the potential to significantly reduce the average wire length of block-to-block interconnects by stacking functional blocks vertically instead of spreading them out horizontally,
b) high-density, high-aspect-ratio connections are possible, which allow implanting complex, multi-chip systems entirely within silicon and
c) RC delays of long, in-plane interconnects are avoided by bringing out-of-plane logic blocks electrically much closer together

The so-called "Inter-Chip-Via (ICV)-SLID concept" [7] is well suited as a chip-to-wafer stacking approach. Starting point are completely processed wafers. Known good dice of the top wafer are aligned bonded to the known good dice of a bottom wafer after wafer-level testing, thinning and separation. This represents the only process step on chip level within the total vertical system integration sequence. The subsequent processing for vertical metallization is on wafer-scale again. Basically, there is no need for additional process steps on stack level.

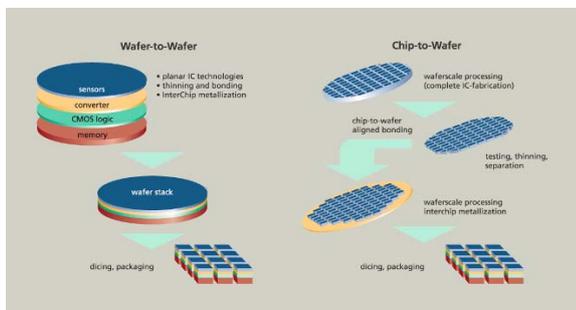

**Fig. 3 VSI concept;-W2W and D2W [7]**

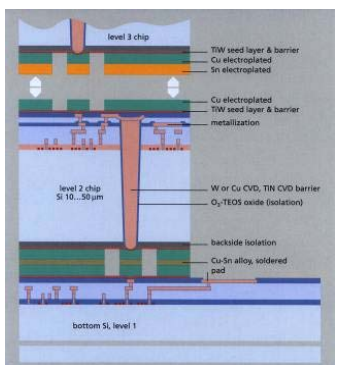

**Fig. 4 Schematic ICV-SLID process**

The ICV-SLID concept is based on the metal-metal bonding of top chips to a bottom wafer by very thin soldering pads (e.g. Cu/Sn) which provide both, the electrical and the mechanical interconnect by solid-liquid-interdiffusion (SLID). The ICV-SLID concept is a non-flip concept. The top surface of the chip to be added is the top surface after stacking it to the substrate. The Through-Si-Vias are fully processed – via formation and metallization – prior to the thinning sequence which has the advantage that the later stacking of the separated known good dice to the bottom device wafer is the final step of the 3D integration process flow. As a fully modular concept, it allows the formation of multiple device stacks. Fig. 4 shows the schematic cross section of a vertically integrated circuit in accordance with the modular "back-to-face" concept, also indicating the stacking of a next level chip. The first essential step of the ICV-SLID process flow is the formation of inter-chip vias. The via etch, lateral isolation and metal filling is performed on wafers with standard thickness, thus resulting in basically high-yield fabrication of inter-chip vias. The ICVs are connected to the contact wiring of the devices by standard metallization (aluminium or copper). The process sequence for the formation of the metalized inter-chip vias is as follows: The ICVs with typically 1-3 µm diameter are prepared on a fully processed and tested device wafer by dry etching (DRIE) through all passivation and multi-level dielectrics layers followed by a deep silicon trench etch. For lateral via isolation, a highly conformal CVD of O3/TEOS-oxide is applied and the inter-chip vias are metalized by using MOCVD of tungsten (MOCVD-TiN as barrier layer) and etched back for metal plug formation. The lateral electrical connection of the Tungsten filled inter-chip vias with the uppermost metal level of the device is performed by standard Al metallization. The devices are now ready for wafer level test and selection. The last process sequence performed on the top wafer with standard thickness is through-mask electroplating of Cu. The top wafer is then temporarily bonded to a handling wafer and thinned with very high uniformity using precision grinding, wet chemical spin etching and a final CMP step until the tungsten-filled vias are exposed from the rear. After deposition of dielectric layers for electrical isolation and opening to the tungsten-filled inter-chip vias, through-resist mask electroplating of Sn/Cu is applied. The surface is completely covered with the soldering metal, electrical contacts are formed by isolation trenches in the Cu/Sn layer and the remaining areas that are not used for electrical means serve as dummy areas for mechanical stabilization of the future stack. The bottom wafer is through-resist mask electroplated with Cu as the counterpart metal of the soldering metal system.

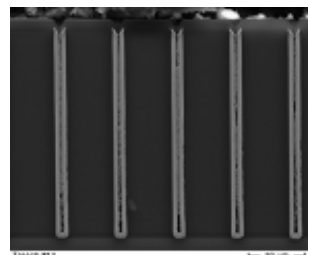

**Fig. 5 High Aspect ratio W filled via**

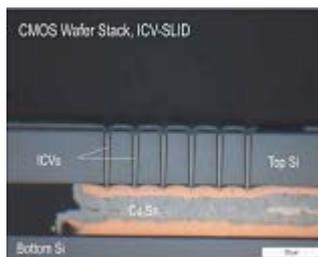

**Fig. 6 Cross-section of interconnect-ted devices with W filled TSV using SLID**

After dicing, the selected known good dice – stabilized with the handling substrates – are picked and placed at the bottom wafer by using a chip-to-wafer bonding equipment with high throughput and an alignment accuracy of 10 µm. The mechanical bond and the electrical contact of the transferred chips are performed in one step by a soldering technology called Solid-Liquid Interdiffusion (SLID). [8]

During the soldering step, at a temperature of approximately 300 °C and applying pressure, the liquid Sn is interdiffused by Cu, finally forming the intermetallic compound (IMC) $Cu_3Sn$. This formed ε- phase is thermodynamically stable with a melting point above 600°C. Using appropriate film thicknesses, tin is consumed and the solidification is completed within a few minutes, leaving copper on both sides. Fig. 6 shows an FIB of a 3-D integrated test structure after soldering and removal of the handling substrate. The tungsten filled ICVs are interconnected by Al wiring to the metallization of the top device and CuSn metal system to the metallization of the bottom device.

## IV. CONCLUSION

Besides the progress in silicon technology following Moore´s law, there is an increasing demand for highly miniaturized complex system architectures. 3D integration based on wafer level approaches has the only potential to meet the requirements of form factor, performance and cost reduction. The ICV stacking concept allows the combination of different devices (e.g. MEMS-sensor, DSP, RF transceiver, power supply). The ICV-SLID technology and micro bump interconnects meet the requirements for device stacking with very high interconnection density ($10^4$-$10^6$ $cm^{-2}$). The combination of silicon interposer with TSV, thin chip integration and VSI opens the way to a new generation of future 3D device architectures.

## V. ACKNOWLEDGEMENT

The authors would like to thank the staff involved in the 3D and Wafer Level System Integration program at Fraunhofer IZM. Special thanks to R. Wieland, Dr. H. Oppermann and K. Zoschke

## VI. REFERENCES

[1] International Technical Roadmap of Semiconductors, TWG A&P, www.itrs.org
[2] F. Binder; "Low Cost Si Carrier – 3D for high density modules", 3D Architecture for Semiconductor Integration and Packaging; San Francisco; Oct.22-24. 2007
[3] M.J. Wolf, P. Ramm, A. Klumpp, "3D-Integration TSV-Technology", EMC 3D Technical Symposium, Munich/Eindhoven, Netherlands, Oct. 2007
[4] M.J. Wolf, P. Ramm, H. Reichl, "3D-System Integration on Wafer Level" SEMI Technology Symposium 2007, International Packaging Strategy Symposium 2007 Semicon Japan, Tokyo
[5] P. Ramm, "3D System Integration: Enabling Technologies and Applications", International Conference SSDM 2006, Yokohama (2006) 318-319
[6] P. Ramm and R. Buchner, "Method of making a vertically integrated circuit", US Patent 5,766,984, Sep. 22, 1994 [DE]
[7] P. Ramm, A. Klumpp, R. Merkel, J. Weber, R. Wieland, "Vertical system integration by using inter-chip vias and solid-liquid-interdiffusion bonding", Japanese Journal of Applied Physics Vol. 43, No. 7A (2004) 829-830
[8] A. Klumpp et.al. "Chip to Wafer Stacking by using Through Silicon Vias and Solid Liquid Interdifusion", 2nd International IEEE Workshop on 3D System integration, Munich(D) Oct 2007
[9] J.H. Lau. "Thermal Stress and Strain in Microelectronic Packaging" Van Nostrand Reinhold, New York, 1993.
[10] B. Wunderle, R. Mrossko, O. Wittler, E. Kaulfersch, P. Ramm, B. Michel, H. Reichl, "Thermo-Mechanical Reliability of 3D-Integrated Microstructures in Stacked Silicon", Mater. Res. Soc. Symp. Proc. 970, MRS 2006 Fall Meeting, Boston, edited by C. A. Bower, P. E. Garrou, P. Ramm, K. Takahashi, Materials Research Society, Warrendale, Pennsylvania (2007) 67-78.
[11] P. Ramm, J.M. Wolf and B. Wunderle. "Wafer-Level 3-D System Integration" in "3-D IC Integration: Technology and Applications", P.E. Garrou, P. Ramm and C.A. Bower, Editors, Wiley-VCH, 2008